\documentclass[12pt, prd, showpacs]{revtex4}

\usepackage{amsfonts}
\usepackage{amssymb}

\begin{document}

\title{Horizons in matter: black hole hair vs. Null Big Bang}

\author{K. A. Bronnikov}
\affiliation{Center for Gravitation and Fundamental Metrology, VNIIMS, 46 Ozyornaya
Street, Moscow 119361, Russia;\\
Institute of Gravitation and Cosmology,
PFUR, 6 Miklukho-Maklaya Street, Moscow 117198, Russia}
\email{kb20@yandex.ru}

\author{Oleg B. Zaslavskii}
\affiliation{Astronomical Institute of Kharkov V.N. Karazin National University, 35
Sumskaya St., Kharkov, 61022, Ukraine}
\email{ozaslav@kharkov.ua}

\date{\today}

\begin{abstract}
\noindent
{\footnotesize {\sl (Essay written for the Gravity Research Foundation
2009 Awards for Essays on Gravitation) }}

\bigskip
  It is shown that only particular kinds of matter (in terms of the
  ``radial'' pressure to density ratio $w$) can coexist with Killing horizons
  in black-hole or cosmological space-times. Thus, for arbitrary (not
  necessarily spherically symmetric) static black holes, admissible are
  vacuum matter ($w=-1$, i.e., the cosmological constant or some its
  generalization) and matter with certain values of $w$ between 0 and $-1$,
  in particular, a gas of disordered cosmic strings ($w=-1/3$). If the
  cosmological evolution starts from a horizon (the so-called Null Big Bang
  scenarios), this horizon can co-exist with vacuum matter and certain kinds
  of phantom matter with $w\geq -3$. It is concluded that normal matter in
  such scenarios is entirely created from vacuum.
\end{abstract}

\keywords{black holes, no-hair theorems, null big bang}
\pacs{04.70.Bw, 04.20.Gz, 04.40 Nr}
\maketitle

  There are well-known issues in gravitation theory which, upon careful
  treatment, exhibit unexpected gaps. In black hole physics, the famous
  no-hair theorems apply to certain kinds of fields (electromagnetic or other
  gauge fields, dilaton scalars etc.) \cite{fn}. Meanwhile, much simpler but
  physically and astrophysically more relevant environments, namely,
  macroscopic matter with certain pressure and density, drop out from
  consideration. Nonetheless, it turns out that if we
\begin{description} \itemsep 1pt
\item [(i)]
  specify the equation of state near the horizon,
\item [(ii)]
  use the Einstein equations and the conservation law and
\item [(iii)]
  employ the horizon regularity condition,
\end{description}
  we obtain that black holes do admit ``matter hair'' but only its very
  specific kinds.

  There is a cosmological counterpart of this issue. Looking what can happen
  beyond a black-hole horizon but very close to it, we can obtain similar
  conditions that restrict possible types of cosmological scenarios starting
  from a horizon, the so-called Null Big Bang (NBB) scenarios. There is some
  complementarity between two seemingly quite different issues, black hole
  hair and NBB scenarios. It is formally expressed in terms of the
  pressure to density ratio $w$: the above two situations are connected by
  the relationship $w \leftrightarrow 1/w$.

  Let us begin with {\bf black holes\/} and consider the general static,
  spherically symmetric metric in the form
\begin{equation}
   ds^2 = A(u)dt^2 -\frac{du^2 }{A^2 (u)}-r^2 (u)(d\theta ^2 +\sin^2 \theta
	\ d\phi^2 )  					\label{m}
\end{equation}
  Horizons of order $n$ correspond to regular zeros of $A(u)$:
  $A \sim (u-u_h)^n$, $n \in {\mathbb{N}}$.

  The stress-energy tensor (SET) of matter can be written as
\begin{equation}
    T_{\mu }^{\nu }={\rm diag}\ (\rho ,-p_{r},-p_{\perp },-p_{\perp }).
							\label{set}
\end{equation}
  Then the combination ${0\choose 0} - {1\choose 1}$ of the Einstain
  equations shows that on the horizon
\begin{equation}  					\label{vac}
	p_{r}(u_{h}) + \rho (u_{h})=0,
\end{equation}
  so the null energy condition (NEC) $T_{\mu }^{\nu }\xi^{\mu}\xi_{\nu}
  \geq 0$, $\xi^{\mu} \xi_{\mu} = 0$ is satisfied on the verge. One can note
  that matter characterized by the condition (\ref{vac}) in the whole space
  is a natural generalization of the cosmological constant in that its SET is
  invariant under radial boosts \cite{dym92}, and, following \cite{bd}, we
  call it vacuum matter. Without losing much generality, we assume that for
  our matter near the horizon $p_{r} = w\rho$, $w = {\rm const}$, $\rho \geq
  0$. If the NEC holds ($w\geq -1$), we call the matter normal, otherwise it
  is said to be phantom. We also admit a noninteracting admixture of vacuum
  matter.

  An analysis according to the above items (i)--(iii) leads to the following
  results \cite{bz08a}.

\medskip\noindent
{\bf Theorem 1}. \ A static, spherically symmetric black hole can be in
  equilibrium with a static matter distribution with the SET (\ref{set}) only
  if near the event horizon either (a) $w\rightarrow -1$ (as in vacuum
  matter) or (b) $w\rightarrow -1/(1+2k)$, $k\in {\mathbb{N}}$.  In case (a),
  the horizon can be of any order $n$ and $\rho (u_{h})\neq 0$.  In case (b),
  the horizon is simple ($n=1)$, and $\rho \sim u-u_{h}$.

\medskip
  The generic case of such hairy black holes is $k=1$ implying $w = -1/3$,
  which, for $p_\bot = p_r$, corresponds to a fluid of disordered cosmic
  strings. Since such strings may be arbitrarily curved or closed, one can
  partly express the meaning of the theorem by the words ``black holes can
  have curly hair''.

  In the presence of vacuum matter, the following theorem holds:

\medskip\noindent
{\bf Theorem 2}. A static, spherically symmetric black hole can be in
  equilibrium with a noninteracting mixture of static nonvacuum matter with
  the SET (2) and vacuum matter with the SET (\ref{set}) only if near the
  event horizon
\begin{equation}  \label{w_bh}
		w\rightarrow -n/(n+2k)
\end{equation}
  where $n$ is the order of the horizon, $n\leq k$, and
  $\rho \sim (u-u_h)^{k}$.

\medskip
  Any amount of other kinds of matter, normal or phantom, added to such a
  configuration, should break its static character by simply falling onto the
  horizon or maybe even destroying the black hole. In other words, black
  holes may be hairy, or dirty, but, in the near-horizon region, normal
  ($p_r \geq 0$) or phantom ($p_r < -\rho$) hair are completely excluded.
  In an equilibrium configuration, all ``dirt'' is washed away from the
  near-horizon region, except vacuumlike or modestly exotic, probably
  ``curly'' hair. In particular, a static black hole cannot live inside a
  star of normal matter with $p_r\geq 0$ unless there is an accretion region
  around the horizon or a layer of string and/or vacuum matter.

  Our approach is also relevant to semiclassical black holes in equilibrium
  with their Hawking radiation (the Hartle-Hawking state), whose SET
  essentially differs from that of a perfect fluid. Since the density of
  quantum fields is, in general, nonzero at the horizon (see Sec. 11 of the
  textbook [1] for details), by the regularity condition (\ref{vac}),  such
  quantum radiation should behave near the horizon like vacuum matter. A
  mixture of Hawking radiation and some kinds of classical matter with
  $-1 < w < 0$ is also admissible.

\medskip\noindent
  {\bf Null Big Bang.}
  Spherically symmetric cosmological models are characterized by the general
  Kantowski-Sachs (KS) metric
\begin{equation}  \label{KS}
    ds^2 =b^2(t) dt^2 -a^2(t) dx^2 -r^2(t)(d\theta^2 +\sin^2\theta d\phi^2).
\end{equation}
  Using the ``quasiglobal'' gauge ($b(t) = 1/a(t)$ and assuming a horizon at
  some $t = t_h$, so that $r(t_h)$ is finite and
\begin{equation}
	a^2 (t)\sim (t-t_{h})^n, \ \ \ n\in \mathbb{N},
\end{equation}
  we can study the near-horizon behavior of the system in the above manner.
  Then the following general results are obtained \cite{bz08b}:
\begin{enumerate}  \itemsep 1pt
\item
  With any normal matter, regular cosmological evolution can only begin
  with a Killing horizon.
\item
  Noninteracting normal matter cannot exist in a KS cosmology near a horizon.
\item
  Normal matter can only appear after a NBB due to interaction with a
  sort of vacuum.
\end{enumerate}

  More specifically, the parameter $w$ should obey the condition $w=-1- 2k/n$
  where $k\geq n >0$, $k, n \in {\mathbb{N}}$. It can be obtained from
  (\ref {w_bh}) by replacing $w\rightarrow 1/w$.

  This generalizes the conclusions made in \cite{bd} for KS cosmologies with
  dustlike matter. As for phantom matter, in its presence a nonsingular
  cosmology is posiible without a NBB. But if there is a horizon, it is
  compatible with phantom matter with $w\leq -3$ only.

  In NBB scenarios, the Universe is highly anisotropic right after crossing
  the horizon, since one of its scale factors, $a(t)$, evolves from zero
  whereas the other, $r(t)$, is finite. So, at this stage, intense particle
  creation from vacuum should occur, leading to rapid isotropization  \cite
  {GMM, bird}. Estimates on a phenomenological level \cite{bd} show that that
  at least some NBB scenarios automatically lead to sufficient isotropy at
  later stages of evolution. This confirms the potential viability of this
  kind of models. Moreover, the existence of a pre-horizon static stage,
  which is observable from the cosmological stage, makes all cosmological
  events causally connected thus removing the inherent difficulties of usual
  Big Bang scenarios, conventionally solved by inflation.

\medskip \noindent
{\bf Abandoning spherical symmetry.}
  It is generally assumed that, in the course of gravitational collapse to a
  black hole, nonspherical perturbatons die out. However, in astrophysical
  conditions (e.g., in double or multiple stellar systems) even static black
  holes, being surrounded by nonsymmetric matter distributions, deflect from
  spherical symmetry. In such more general black holes, generic properties of
  Killing horizons should play an important role in a (not yet quite
  well understood) relationship between the symmetry of horizons and black
  hole entropy (see, e.g., \cite{vis} and references therein).

  It turns out, however \cite{bz09}, that almost all restrictions on the values 
  of the parameter $w$ compatible with horizons, described above, hold in the
  general static case. Such results are obtained with the above items
  (i)--(iii), now applied to a general static metric written with the aid of
  the Gaussian coordinate $l$, orthogonal to a family of 2-surfaces,
  one of which ($l=l_h$) corresponds to the horizon:
\begin{equation}
    ds^{2}=-N^2 dt^2 +dl^2 + \gamma_{ab}dx^{a}dx^{b}, \ \ \ \ a,b=2,3.
							\label{n}
\end{equation}
  where $N = N(l, x^2, x^3)$ and $\gamma_{ab}= \gamma_{ab}(l,x^2,x^3)$.
  This metric was used in \cite{vis} in discussing regularity conditions
  on the horizon, and the relation (\ref{vac}) was obtained, where
  $p_r$ now denotes pressure in the $l$ direction).

  As in spherically symmetric configurations, the possible existence or
  non-existence of a black hole inside matter and the horizon type depend on
  the value of $w$ and on the presence or absence of vacuum matter with the
  density $\rho_{\rm vac}$, and analogs of Theorems 1 and 2 are valid.
  Nevertheless, some new features do appear \cite{bz09}. Thus, first, in the 
  absence of vacuum matter, degenerate horizons ($n \geq 2$) are now possible 
  if the corresponding 2-metric $\gamma_{ab}$ is flat, which is true, e.g., for
  cylindrical black holes. Second, if $\rho_{\rm vac}\neq 0$, the sign of
  $\rho_{\rm vac}(l_h)$ coincides with that of two-dimensional curvature of the
  horizon. In particular, $\rho_{\rm vac} < 0$, forbidden in the spherical
  case, is compatible with a horizon with hyperbolic geometry. Thus matter
  with certain negative values of $w$ can contain horizons with non-spherical 
  topologies (cf. \cite{un}).

  Horizons in nonsymmetric NBB cosmologies can be considered in a similar
  way, with almost the same results as described here, to be presented
  in detail elsewhere. It would be of interest to obtain explicit examples of
  non-KS nonsingular NBB scenarios. In the KS case, such examples are already
  known, one with phenomenological vacuum matter \cite{bd} and another with
  a phantom scalar field \cite{pha1}.

\medskip
  To conclude, in our reasoning, which has been entirely local and relied on
  near-horizon expansions, we did not assume any particular equation of state
  for matter and even did not restrict the behaviour of the transverse
  pressure except for its regularity requirement. In this sense, our
  conclusions are model-independent.

\end{document}